# Game theory in Oligopoly

Prof. Marx Boopathi, Nikolaj Sujen.

Abstract

The game theory techniques are used to find the equilibrium of a market. Game theory refers to the ways in which strategic interactions *among* economic agents *produce* outcomes *with respect to the* preferences *(or* utilities*)* of those agents, where the outcomes in question might have been intended by none of the agents. The oligopolistic market structures are taken and how game theory applies to them is explained.

1. Introduction

The market structure identifies how a market works in terms of the number of the firms engaged in, the nature of the produced product, efficiency of firm etc. There are many forms of market structure and the most revealed is Oligopoly. In oligopolistic markets, there always remain a struggle between self interest and cooperation. If the output is limited by all the firms, the price becomes high, but it implies that then output will be expanded because of firm's incentive to do that.

The analysis through Game theory is directly relevant to the behaviour of businesses in oligopolistic markets. The decisions taken using the theory about the self interest and cooperation decide the investment and spending of the firms. Conventional economics describes the operation of mature & established markets, but it throws no light on people's creativity in finding new ways of interacting with one another. The economics of markets is dynamic and evolving.

The innovation continues and nothing is taken given. In the rapidly and free-form transforming marketplace, game theory is the kernel of a new economics.

2. The Oligopoly

The main characteristics of an oligopoly market are:

- There are few firms to compete in. It may happen that a large number of firms are there but two very big manufacturers dominate the industry.
- The Concentration Ratio of the market sales is in the hand of those two producers. If they account for 90% ratio of share of market sales in the industry it is called a 2 firm concentration ratio of 90%. The airlines industry is a good example of it.
- The larger of the two firms has Price leadership that is followed by smaller firm
- Collusion Potential
- Interdependence between firms—Behaviour is affected by actions of rivals
- Goods manufactured may be either highly differentiated or homogeneous
- loyalty against brand & branding may cause a potential source of advantage
- Cournot Model suggest equilibrium in long run cause normal profits and equal market share
- Use of game theory to explain behaviour of the firms to a extent

- AC curve is saucer shaped – minimum efficiency may be needed to cover large range of output
- Entry restricted due to high barriers
- Examples include Moody's & S&P in the ratings market, Coca cola and Pepsi, Visa and Master card etc.

3. Game theory

"It is a universal language for the unification of the behavioural sciences." Gintis (2000, 2009).

In simple words this theory studies peoples' behaviour in strategic situations. It is the formal study of cooperation and conflict and can be applied where the actions of several firms are interdependent. The game theory concepts provide a language to structure, formulate, analyze and understand strategic scenarios.

Edward C Rosenthal, (2011, p.18) states that "Game Theory is the study of the ways in which strategic interactions among rational players produce outcomes with respect to the preferences of those players, none of which might have been intended by any of them."

In a duopoly or the oligopolistic market, each firms' profits depend on other firms' actions resulting in the "prisoners' dilemma". The prisoners' dilemma is a particular game, which illustrates the difficulty of cooperation, even when it is in the best interests to market players. The self owned dominant strategies are selected by the both market players for their short-sighted personal profits. Eventually, equilibrium is reached where both may worse off than they would have been, if an alternative (non-dominant) strategy could have been agreed upon between them.

The fundamental assumption of the game theory is that all players are intelligent and rational and each player uses those strategies which result in the long run equilibrium. At this point not anyone wants to withdraw from its strategic ideology.

In the grocery market's duopolistic competition, like in the Australian case, the game theory is particularly important as it offers Low barriers to entry without any license, any upfront cost or any infrastructure. Because of the already present two big bulls, most retailers avoid entering into market because they can not afford.

3.1 Nash Equilibrium

In the supermarkets duopolistic competition, like in the Australian case, the strategies adopted are related to EDLP (everyday low pricing) and HLP (high low pricing) that change from week to week. Both strategies are dominant for maximising sales and profits with focus on week to week sales. The strategy related to HLP is more transparent and explicit, which consists of in store advertising, media advertising, merchandising and price reductions. This strategy in game theory is shown in the following table. As per the game theory it is a normal game between the two players.

| | Price Reduction ($\alpha_i$) | Merchandising ($\beta_i$) In-Store Advertising | ISI R.TV, and Ph..." ($\gamma_i$) Media Advertising— |
|---|---|---|---|
| | $(\alpha_i, \alpha_j)$ $(\alpha_i > \alpha_j)$ or $(\alpha_i \leq \alpha_j)$ | $(\alpha_i, \beta_j)$ $(\alpha_i > \beta_j)$ or $(\alpha_i \leq \beta_j)$ | $(\alpha_i, \gamma_j)$ $(\alpha_i > \gamma_j)$ or $(\alpha_i \leq \gamma_j)$ |
| | $(\beta_i, \alpha_j)$ $(\beta_i > \alpha_j)$ or $(\beta_i \leq \alpha_j)$ | $(\beta_i, \beta_j)$ $(\beta_i > \beta_j)$ or $(\beta_i \leq \beta_j)$ | $(\beta_i, \gamma_j)$ $(\beta_i > \gamma_j)$ or $(\beta_i \leq \gamma_j)$ |
| | $(\gamma_i, \alpha_j)$ $(\gamma_i > \alpha_j)$ or $(\gamma_i \leq \alpha_j)$ | $(\gamma_i, \beta_j)$ $(\gamma_i > \beta_j)$ or $(\gamma_i \leq \beta_j)$ | $(\gamma_i, \gamma_j)$ $(\gamma_i > \gamma_j)$ or $(\gamma_i \leq \gamma_j)$ |

Table 1: A Modified Illustration of a Normal-form Game.
Note that the cells do not show symmetrical levels of strategies but simply all show strategy that is possible. PR is Price-Reduction, M is Merchandising and MA is Media-Advertising.

Source: An application of Game theory

The parameters in above table can take any values but the constraint is that each player has played the game several times and knows other players components of strategy which may be different. That means each player has in all nine strategies and each play requires three strategies at one time. Thus in all there are twenty seven pay offs for this game. This implies that it would not be easy to achieve Nash equilibrium but not impossible.

The sales being observed on weekly basis, if these are in line with expectations and long run objectives of player one and if rivals are following passive strategies then the probability that player two too will continue the passive strategy is quite high and vice versa. The same strategy is applied for aggressive strategies also. However, within this game the costs are likely to remain low and profits may soar.

In practice, however it does not happen and rivals may adopt an alternate strategy to be better off. The following table shows the pay off matrix for nine of twenty seven strategies that are implicit in the above table.

|  | | Player j | |
|---|---|---|---|
|  | TPE$_A$ | TPE$_M$ | TPE$_P$ |
| Player i TPE$_A$ | $\pi_{A,A}$ (13, 13) | $\pi_{A,M}$ (16, 12) | $\pi_{A,P}$ (30, 8) |
| TPE$_M$ | $\pi_{M,A}$ (12, 16) | $\pi_{M,M}$ (14, 14) | $\pi_{M,P}$ (25, 12) |
| TPE$_P$ | $\pi_{P,A}$ (8, 30) | $\pi_{P,M}$ (12, 25) | $\pi_{P,P}$ (20, 20) |

Source: An application of Game theory

The each cell in the above table is showing the pay off for each player for all the three different strategies. The Nash equilibrium is also shown but this level also can't be considered as a stable equilibrium because it is attributed to higher inventory and promotional costs.

If instead of two, there are more firms in the market, the price effect will be smaller resulting in the lower Nash equilibrium price. If the number of firms reaches infinity, the price effect will also approach zero. So, output will be increased by each seller whenever the price moves above MC. In the end, the perfectly competitive price will be there and the quantity would also be socially efficient.

M. Shubik, (1968,p.) claims that

The Nash equilibrium in the general game does not yield a higher payoff, given the other players chosen strategies. It is a notion of joint rationality as each player best responds to the other players.

3.2 Bertrand Equilibrium

*For differentiated products, two identical firms choose prices simultaneously and end up with the situation likely to perfect competition.*

*The only pure-strategy Nash equilibrium is $p_1^* = p_2^* = c$ because the best response is played by both firms to each other*

*Neither firm has an incentive to deviate to some other strategy.*

*If $p_1$ and $p_2 > c$, a firm could gain by undercutting the price of the other and capturing all the market. If $p_1$ and $p_2 < c$, profit would be negative*

To hack with quantity and profits, if both firms compete viciously on price then equilibrium will occur only where price equals the marginal cost. This is because both firms are producing identical products and if one reduces its price the other will not be left behind and try to cut its price still lower. However this equilibrium is socially efficient as consumers will be benefitted most.

3.3 Cournot Equilibrium

If both firms become price takers and put the price as secondary option and consider quantity produced they can achieve Cournot equilibrium by applying game theory. As the product is identical profit maximisation using quantity for one firm depends on the quantity produced by the other one. In this model

- *firms set quantities rather than prices and each firm's own decisions are recognised about $q_i$ affect price*
  - $\partial P/\partial q_i \neq 0$
- *However, each firm thinks that any other firm is not affected by its decisions*
  - $\partial q_j/\partial q_i = 0$ for all $j \neq i$
- *For profit maximization, the FOC are*
  $$\frac{\partial \pi_i}{\partial q_i} = P(Q) + P'(Q)q_i - C'_i(q_i) = 0$$
- The firm maximizes profit where $MR_i = MC_i$
- *Price exceeds marginal cost by* $P'(Q)q_i$

If q1 is the qty produced by one firm and q2 is the qty produced by other firm, then the total qty produced will be q=q1+q2. Using game theory, the solution will be

q1=q2= (a-cb)/3, where c is the MC of the firm

It can be noted that this equilibrium is in between monopoly and oligopoly production levels and the profit margins earned by the two firms are less than half of those as earned by the monopolist.

4. An example

For example, a few, large firms dominate many Australian markets. A duopoly exists in the grocery market of Australia, where it is dominated by Coles and Woolworths, nationally. Both of these have extensive and wide distribution systems, with many stores in almost all in regional and urban areas across Australia. Other firms, if they wish to enter need a large investment to compete with these.

In the grocery market of Australia, the seller concentration ratio is approx 80%; the two largest firms Coles and Woolworths account for 80% of total sales of groceries in Australia. In a study of retail marketing Bovil (2008) stated that "When you have that much (market share) you create market distortion - a supplier who has to deal with someone with that much power works from a weak position. Two retailers dominating sales did not qualify as a "market" under economic definitions."

In an oligopolistic market competition is intense. Both non-price and price techniques are used to compete. The warranties and tricky advertisements are common aspect. As per an article in *weekly times*, marketing manager, Woolworths (2001),

We will not be beaten on price. If our competitor is selling item X for less, we will match their prices, whether the item's price has been reduced for a special sale or not. We will replace the item without cost if it is found to be defective in any way.

In particular, Woolworth's has succeeded with its "We're the fresh food people" advertising campaign.

In oligopoly mutually interdependence is a must. If policy of price change is adopted by one, then this will affect the sales of other firm in the market. This can be understood as follows.

Suppose Coles and Woolworths initially sell tomatoes @ $6 per KG. Woolworths lowers its 2price by $1 per KG so that its sales rise from 10,000 KG per week to 20,000 KG per week. This implies that at this stage demand curve is elastic for tomatoes; the 17% decrease in price has resulted in 50% increase in volume of sales. So, as Coles begins to lose its sales to Woolworth's, to match Woolworths price, it also reduces price by $1 per KG.

Coles does not stop here and decides to go further by reducing its prices to $3 per KG. The reason provided that when Woolworths decreased its price, it made less profit per kilogram, but as many more kilograms of tomatoes were sold, overall it made greater profits. If Coles do the same, it will end up making greater profits, too.

However, practically it won't happen. With Cole's new price, its sales increase from 20,000 KG per week to 25,000 KG per week.

The reason behind this is the kinked demand curves which many oligopolies face in practice, which has two parts; one an inelastic part, and the other one is elastic.

In Woolworths case the demand was elastic but in Coles case it turned out to be inelastic. With falling prices, greater amounts can not be bought forever. The buying patterns will switch from Woolworths to Coles. However as both are close substitutes, price changes will result in marked reduced volume of sales, from one to the other and the total market demand for tomatoes becomes inelastic as the price is already low.

In the following figure of kinked demand curve it can be seen that above the kink, the demand of tomatoes is elastic as Coles' prices does not changed. But below the kink, it becomes relatively inelastic as now Coles has introduced a similar reduction in price, ultimately leading to a price war between the two. Thus, the best choice for both will be to produce at point **E,** as that is the kink point at which equilibrium occurs.

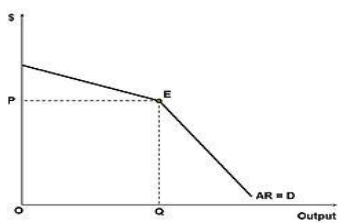

Figure 1: Kinked demand curve

Woolworths knows that with its price reduction, it will make more sales, but only if price reduction step is not taken by Coles in response. If Coles does and if Woolworths also cut its price even further, in overall no more profits will be gained by either of both.

The profits will come to very low levels but Consumers will get benefited.

5. Conclusion

It can be said that in oligopoly, the firms have their concentration in non competitive non price areas; such as after-sales service and advertisement. The firms try to make differentiation in their products in the eyes of consumers, which can be done in many ways like making improved quality products, different wrapping or packaging, bonus or scheme offerings etc. These are some ideas that give more opportunities to each firm to be different from its rivals in terms of output and prices. The firms of oligopolistic markets may not be interested in real competition, benefitting consumers with lower prices. Rather, they operate in ways to maintain their "cosy" share of the market and high levels of profits. This is the reason why most of the time prices tend to stay steady in oligopoly.